\documentclass[12pt]{article}

\usepackage{amsfonts,amsmath,amssymb,accents,mathrsfs}
\usepackage[retainorgcmds]{IEEEtrantools}
\usepackage{multirow}
\usepackage{multicol}
\usepackage{tikz}
\usepackage{graphicx,color}
\usepackage{booktabs}
\usepackage{placeins}
\usepackage[colorlinks,linkcolor=Blue,citecolor=Blue,urlcolor=Blue,bookmarks,
bookmarksnumbered]{hyperref}
\usepackage[nosort]{cite}
\usepackage[scaled=0.85]{helvet}
\usepackage[T1]{fontenc}
\usepackage[utf8]{inputenc}
\usepackage{cancel}
\definecolor{Red}    {rgb}{0.90,0.00,0.12} %  1
\definecolor{Blue}   {rgb}{0.00,0.00,1.00} %  2
\definecolor{Green}  {rgb}{0.10,0.70,0.10} %  3
\definecolor{Turque} {rgb}{0.00,0.65,0.85} %  4
\definecolor{Orange} {rgb}{1.00,0.50,0.15} %  5
\definecolor{Magenta}{rgb}{1.00,0.00,1.00} %  6
\definecolor{Gold}   {rgb}{1.00,0.75,0.25} %  7
\definecolor{Seaweed}{rgb}{0.01,0.24,0.09} %  8
\definecolor{Purple} {rgb}{0.50,0.25,0.55} %  9
\definecolor{Brown}  {rgb}{0.43,0.26,0.32} % 10
\definecolor{grey1}  {rgb}{0.20,0.20,0.20} % 11
\definecolor{grey2}  {rgb}{0.40,0.40,0.40} % 12
\definecolor{grey3}  {rgb}{0.60,0.60,0.60} % 13
\definecolor{grey4}  {rgb}{0.80,0.80,0.80} % 14
\definecolor{grey5}  {rgb}{0.90,0.90,0.90} % 15

\def\a{{\alpha}}
\def\b{{\beta}}

\def\s{{\sigma}}
\def\S{{\Sigma}}
\def\t{{\tau}}

\def\th{{\theta}}

\def\ad{{\dot{\alpha}}}
\def\bd{{\dot{\beta}}}

\def\thd{{\bar{\theta}}}

\def\N{{\mathcal{N}}}

\def\D{{\rm D}}
\def\Dd{{\bar{\rm D}}}
\def\pa{\partial}

\def\be{\begin{equation}}
\def\ee{\end{equation}}
\def\bea{\begin{IEEEeqnarray*}}
\def\eea{\end{IEEEeqnarray*}}
\def\n{\IEEEyesnumber}

\makeatletter
\def\section{\@startsection{section}{1}{\z@}
              {3ex plus-1ex minus-.2ex}{1pt plus1pt}
              {\large\sf\bfseries\boldmath}}
\def\subsection{\@startsection{subsection}{2}{\z@}
              {1.5ex plus-1ex minus-.2ex}{0.01pt plus1pt}{\sf\slshape}}
\def\subsubsection{\@startsection{subsubsection}{3}{\z@}
              {1.5ex plus-1ex minus-.2ex}{0.01pt plus0.2pt}{\sf\boldmath}}
\def\paragraph{\@startsection{paragraph}{4}{\z@}
              {.75ex \@plus.5ex \@minus.2ex}{-2mm}{\sf\bfseries\boldmath}}
\makeatother

   \parskip\medskipamount     \lineskip=0pt
\thicklines
\begin{document}
\thispagestyle{empty}
\noindent{\small
\hfill{HET-1787 {~} \\ % un-comment-out and specify when done}
$~~~~~~~~~~~~~~~~~~~~~~~~~~~~~~~~~~~~~~~~~~~~~~~~~~~~~~~~~~~~$
$~~~~~~~~~~~~~~~~~~~~\,~~~~~~~~~~~~~~~~~~~~~~~~~\,~~~~~~~~~~~~~~~~$
 {~}
}
\vspace*{6mm}
\begin{center}
{\large \bf 
Superfield continuous spin equations of motion
\vspace{3ex}
} \\   [9mm] {\large { I. L.
Buchbinder,\footnote{joseph@tspu.edu.ru}$^{a,b}$ S.\ James Gates,
Jr.,\footnote{sylvester\_gates@brown.edu}$^{c}$ and K.\
Koutrolikos\footnote{konstantinos\_koutrolikos@brown.edu}$^{c}$ }}
\\*[8mm]
\emph{
\centering
$^a$Department of Theoretical Physics,Tomsk State Pedagogical University,\\
Tomsk 634041, Russia
\\[6pt]
$^b$National Research Tomsk State University,\\
Tomsk 634050, Russia
\\[6pt]
$^{c}$Department of Physics, Brown University,
\\[1pt]
Box 1843, 182 Hope Street, Barus \& Holley 545,
Providence, RI 02912, USA
}
 $$~~$$
  $$~~$$
 \\*[-8mm]
{ ABSTRACT}\\[4mm]
\parbox{142mm}{\parindent=2pc\indent\baselineskip=14pt plus1pt
We propose a description of %manifestly
supersymmetric continuous
spin representations in $4D,\N=1$ Minkowski superspace at the level
of equations of motions. The usual continuous spin wave function
is promoted to a chiral or a complex linear superfield which includes both 
the single-valued
(span integer helicities) and the double-valued (span half-integer
helicities) representations thus making their connection under
supersymmetry manifest. The set of proposed superspace constraints for
both superfield generate the expected Wigner's conditions for both 
representations.  
}
\end{center}
$$~~$$
\vfill
\vfill
\clearpage
%
%%%%%%%%%%%%%%%%%%%%%%%%%%%%%%%%%%%%%%%%%%%%%%
\section{Introduction}
\label{intro}
%%%%%%%%%%%%%%%%%%%%%%%%%%%%%%%%%%%%%%%%%%%%%%
It is generally accepted that the elementary excitation in any
fundamental theory are classified according to the symmetries of the
vacuum. It follows that the free elementary particles are associated
with the unitary and irreducible representations of relevant
spacetime symmetry groups. For this reason, special attention is
paid to the consideration of maximally symmetric spacetimes
(Minkowski, de Sitter and anti-de Sitter).  In the case of $4D$
Minkowski spacetime, Wigner \cite{W}, classified all such
representations. The one particle representations are 
labeled by the mass and spin quantum numbers, which correspond to the 
eigenvalues of the two invariant
Casimir operators (quadratic) $\mathcal{C}_1=P^mP_m$ and (quartic) 
$\mathcal{C}_2=W^mW_m$, where $P_m$
and $W_m$ are the momentum and Pauli-Lubanski vectors respectively. The 
one particle states inside the representations are labeled by the 
eigenvalues of the corresponding Cartan subalgebra (like the spin/helicity 
in the direction of motion).

%Some of these representations appear in local field theories and string 
%theories. These are the familiar finite size representations that describe 
%massless particles with fixed integer or half-integer helicity and massive 
%particles with integer or half-integer spin. A subset of them have been 
%observed in nature. Other representations are the tachyonic particles 
%which are characterized by negative eigenvalues of $\mathcal{C}_1$. Their 
%presence indicates instabilities and they do not appear in nature. The 
%rest, fall in the category of \emph{continuous spin} representations (CSR)
%\cite{csr1,csr2,csr3,csr4,csr5,csr6,csr7,csr8,csr9,csr10,csr11}.
%This type 
%of representations are massless (vanishing $\mathcal{C}_1$) and are 
%characterized by a nonvanishing eignevalue of the second Casimir 
%$\mathcal{C}_2=\mu^2$, where $\mu$ is a real, continuous
%parameter with dimensions of mass. There are two such representations, the
%signle-valued one and the double-valued. The size of both of them is 
%countable infinite and their spectrum includes all integer separated 
%integer or half integer helicity states respectively with multiplicity 
%one\footnote{An alternative terminology for such representations is 
%``infinite spin''. This is a more appropriate name because it captures the 
%essence of the spectrum of these representations (spin is not bounded)
%and avoids the use of the misleading term ``continuous spin'' (the 
%helicity of the states is not continuous). Nevertheless, for historical 
%reasons continuous spin is the prevailed nomenclature and that is
%the one we will use.}.

Some of these representations appear in local field theories and
string theories. These are the familiar finite size representations
that describe massless particles with fixed integer or half-integer
helicity and massive particles with integer or half-integer spin. A
subset of them have been observed in nature. Other representations
are the tachyonic particles which are characterized by negative
eigenvalues of $\mathcal{C}_1$. Their presence indicates
instabilities and they were never observed. The rest, fall in the
category of \emph{continuous spin} representations (CSR)
\cite{csr1,csr2,csr3,csr4,csr5,csr6,csr7,csr8,csr9,csr10,csr11}.
This type of representations are massless (vanishing
$\mathcal{C}_1$) and are characterized by a non vanishing eigenvalue
of the second Casimir $\mathcal{C}_2=\mu^2$, where $\mu$ is a real,
continuous parameter with dimensions of mass. There are two such
representations, the single-valued one and the double-valued. The
size of both of them is countable infinite and their spectrum
includes all integer separated integer or half integer helicity
states respectively with multiplicity one\footnote{An alternative
terminology for such representations is ``infinite spin''. This is a
more appropriate name because it captures the essence of the
spectrum of these representations (spin is not bounded) and avoids
the use of the misleading term ``continuous spin'' (the helicity of
the states is not continuous). Nevertheless, for historical reasons
continuous spin is the prevailed nomenclature and that is the one we
will use.}.

The infinite number of degrees of freedom per spacetime point was the 
reason why Wigner rejected the use of such representations, claiming that
the heat capacity of a gas of continuous spin particles is infinite 
\cite{W2}. Further attempts to relate these representations with physical  
systems have also failed. In ref. \cite{csr5,csr6} it was shown that the 
free field description of these representations breaks causality or 
locality thus making impossible to construct a consistent
quantum field theoretic description. Not surprising, these representations 
have been ignored. Yet, the same two exotic properties (presence of a 
continuous dimensionfull parameter and an infinite tower of massless 
helicities) are very appealing from the point of view of higher spin 
(gravity) theories. Consistent interacting higher spin theories require 
the presence of infinite massless particles with arbitrary
high helicities \cite{V} and a dimensionfull parameter to weight the 
higher number of derivatives required by the interactions. This 
dimensionfull parameter is usually identified with the radius of (A)dS 
spacetime. CSR naturally provide both these features, hence in principle 
it can be seen as a good candidate model for interacting higher spins in 
flat spacetime and even possibly bypassing some of the no-go theorems 
\cite{csr11}. For this reason, recently there has been
an increased interest in the study of CSR 
\cite{rcsr1,rcsr2,rcsr3,rcsr4,rcsr5,rcsr6,
rcsr7,rcsr8,rcsr9,rcsr10,rcsr11,rcsr12,rcsr13,rcsr14,rcsr15,rcsr16,
rcsr17,rcsr18,rcsr19,rcsr20,rcsr21,rcsr22} investigating various kinematic
(covariant on-shell, off-shell descriptions) and dynamic (interactions, 
scattering amplitudes) properties (for a review see \cite{rcsr13}).

The object of this work is to study the continuous spin representations in 
the presence of supersymmetry. In ref. \cite{csr7} it was shown that the 
supersymmetry charges are compatible (commute) with the transverse vector 
generators of $iso(3,1)$: $\Pi^i$, $i=1,2$
which define the CSR. Hence, it is expected that the single-valued CSR can 
be combined with the doubled-valued CSR in order to assemble 
supersymmetric continuous spin representation (sCSR) that include both 
integer and half-integer helicities. In this letter we aim to find a
covariant, on-shell description of sCSR that makes the supersymmetry 
manifest. For that we use the $4D$ Minkowski, $\mathcal{N}=1$ superspace 
formulation. We find that the usual Wigner wavefunction is elevated to a 
chiral or complex linear superfield. It's bosonic component correspond
to two single valued CSR and its fermionic component corresponds to one 
double-valued CSR. We propose a set of covariant superspace constraints 
that such a wavefunction must satisfy on-shell in order to describe a sCSR 
as this is defined by the super-Poincar\'{e} algebra. By projecting into 
components we show that these constraints give back Wigner's equations for 
the single and double valued CSR as expected.  During the time
of writing, new work appeared \cite{rcsr23} which demonstrates the 
connection of the 
single valued CSR with the double valued CSR under on-shell supersymmetry.
The authors showed that on-shell supersymmetry transformations map
one CSR to the other. Our results generalize this for off-shell 
supersymmetry transformation since the superfields provide the necessary 
auxiliary fields to close the algebra of the transformations without the use of equations.

The plan of the paper is as follows. In section 2, we review the group 
theoretical description of CSR using the method of the 
eigenvalues of the Casimirs of the Poincar\'{e} algebra.
Then we review the discussion of massless representations of the super-
Poincar\'{e} algebra and we extend the discussion to the definition of
sCSR. In section 3, we consider the superfield description of such 
representations and derive the superspace, covariant, equations of motion 
it must satisfy. In section 4, we discuss the component projection and 
recover the on-shell description of the single valued and
double valued CSR.
%
%%%%%%%%%%%%%%%%%%%%%%%%%%%%%%%%%%%%%%%%%%%%%%
\section{Review of CSR and sCSR from the symmetry algebra viewpoint}
\label{sec2}
%%%%%%%%%%%%%%%%%%%%%%%%%%%%%%%%%%%%%%%%%%%%%%
For the definition and classification of the various representations, we 
are following the method of diagonalizing the Casimirs and the Cartan 
subalgebra generators of the stabilizer subgroup
of the four dimensional Poincar\'{e} group and its $\mathcal{N}=1$ 
supersymmetric extension.
\subsection[4D Poincare algebra]{$4D$ Poincar\'{e} algebra}
The Poincar\'{e} group\footnote{Its part connected to the identity} is the 
group of isometries of Minkowski spacetime and it is generated by the set 
of rigid motions, i.e. translations ($\mathbb{P}_m$) and rotations ($
\mathbb{J}_{mn}$) which satisfy the algebra
\footnote{We follow the discussions in \cite{csr2,BB,rcsr4}}
:
\bea{l}\n\label{Poincare}
{}[\mathbb{J}_{mn},\mathbb{J}_{rs}]=i\eta_{mr}\mathbb{J}_{ns}-i\eta_{ms}
\mathbb{J}_{nr}+i\eta_{ns}\mathbb{J}_{mr}-i\eta_{nr}\mathbb{J}_{ms}~,\\
{}[\mathbb{J}_{mn},\mathbb{P}_{r}]=i\eta_{mr}\mathbb{P}_{n}-i\eta_{nr}
\mathbb{P}_{m}~,\\
{}[\mathbb{P}_m,\mathbb{P}_n]=0~.
\eea
The stabilizer subgroup (little group) is generated by
the translations generator $\mathbb{P}_m$ and the Pauli-Lubanski vector 
$\mathbb{W}_m$,~
$\mathbb{W}^m=\tfrac{1}{2}\varepsilon^{mnrs}\mathbb{J}_{nr}\mathbb{P}_{s}$ 
with the algebra
\bea{l}\n\label{WW}
[\mathbb{W}_m,\mathbb{P}_n]=0~~,~~
[\mathbb{W}^m,\mathbb{W}^n]=i\varepsilon^{mnrs}\mathbb{W}_r\mathbb{P}_s
~~,~~\mathbb{W}^m\mathbb{P}_m=0~,\\
{}[\mathbb{J}_{mn},\mathbb{W}_{r}]=i\eta_{mr}\mathbb{W}_{n}
-i\eta_{nr}\mathbb{W}_{m}~.
\eea
It is useful to keep in mind that when we consider for realizations of 
representations of the above algebra in terms of fields (wavefunctions)
the abstract generators $\mathbb{J}_{mn},~\mathbb{P}_m$ are replaced by 
the corresponding operators
$\mathcal{J}_{mn}=-ix_{[m}\pa_{n]}-i\mathcal{M}_{mn},~
\mathcal{P}_m=-i\pa_{m}$ 
that generate the group action in the space of fields. Notice that 
$\mathbb{W}_m$, due to its structure, does not depend on the ``orbital'' 
part ($x_{[m}\mathbb{P}_{n]}$) of $\mathbb{J}_{mn}$ and only the 
``intrinsic'' spin ($\mathcal{M}_{mn}$) part survives. The Casimirs of 
(\ref{Poincare}) are $\mathcal{C}_1=\mathbb{P}^2$ and
$\mathcal{C}_2=\mathbb{W}^2$ and their eigenvalues label the 
representations. Additionally, $\mathbb{P}_m$ is one of the Cartan s
subalgebra generators hence the one particle states have at least one 
additional label $p_m$, the eigenvalue of $\mathbb{P}_m$. The eigenvalue 
(label) $p_m$ is restricted to the finite set of representative momenta 
$k_m$, since all other possible momenta are generated by the action of 
various group elements (boosts). The different kinds of representative 
momenta are that of a massive particle at rest $k_m=(-m,0,0,0)$ 
(timelike), that of a massless particle $k_m=(-E,0,0,E)$ (lightlike) or 
that of a tachyonic particle $k_m=(m,0,0,0)$ (spacelike).

For massless ($\mathcal{C}_1=0$) particles, $\mathbb{P}_{m}$ will be 
identified with $p_m=(-E,0,0,E)$. The orthogonality of $\mathbb{W}_m$
with $\mathbb{P}_m$ fixes the structure to the Pauli-Lubanski vector to 
be:
\bea{l}\n\label{W}
\mathbb{W}_m=p_m W+\Pi_{m}%(-W,\Pi_1,\Pi_2,W)
\eea 
where $\Pi_{m}$ is the transverse vector $\Pi_m=(0,\Pi_1,\Pi_2,0)$ with 
$\Pi_1=E(\mathbb{J}_{23}+\mathbb{J}_{20}),~\Pi_2=-E(\mathbb{J}_{13}+
\mathbb{J}_{10})$ and $W=\mathbb{J}_{12}$. According to (\ref{WW}) these 
elements satisfy the algebra
\bea{l}\n\label{Pi}
[\Pi_1,\Pi_2]=0~~,~~i[\Pi_1,W]=\Pi_2~~,~~i[\Pi_2,W]=-\Pi_1
\eea
This is the algebra of group $E2$ and describes the symmetries of the two 
dimensional euclidean plane perpendicular to the motion of the massless 
particle\footnote{$\Pi_i$ are the two generators of translations along the 
two perpendicular directions and $\mathbb{J}_{12}$ is the generator of 
rotation along the axis of motion}. Instead of $\Pi_i$ one can use the 
equivalent set of generator $\Pi^{\pm}=\Pi_1\pm i \Pi_2$, where the above 
algebra takes the simpler form:
\bea{l}\n\label{Pipm}
[\Pi^{\pm},\Pi^{\mp}]=0~~,~~[\mathbb{J}_{12},\Pi^{\pm}]=\pm~\Pi^{\pm}
\eea
The second Casimir takes the form
\bea{l}\n
\mathcal{C}_2=(\Pi_1)^2+(\Pi_2)^2=\Pi^{+}~\Pi^{-}
\eea
Due to the structure of (\ref{Pi}) there are two natural sets of 
eigenstates that one can use in order to describe the various 
representations. The first set includes the helicity states 
($|\lambda\rangle$), which are the eigenstates of $\mathbb{J}_{12}$.
They are labeled by a discrete integer or half-integer helicity
and they are mixed under the nontrivial action of $\Pi_i$. The second set 
includes the angle states, which are the eigenstates of $\Pi_1$ and $
\Pi_2$. They are labeled by a continuous angle\footnote{This is the origin
of the ``continuous'' spin terminology.} parameter and the action of 
$\mathbb{J}_{12}$ results to a shift of this angle. The two sets of states
are related through a Fourier transformation. The definition of helicity 
states is:
\bea{l}\n\label{hel}
\mathbb{J}_{12}|\lambda\rangle=\lambda|\lambda\rangle~~,~~
\mathcal{C}_2|\lambda\rangle=\mu^2|\lambda\rangle
\eea
where the eigenvalue of $\mathcal{C}_2$ is a real positive number 
parametrized by the dimensionfull parameter $\mu$. It is straight forward 
to show that the action of $\Pi^{\pm}$ on a helicity 
state increases (or reduces) the helicity by one unit
\bea{l}\n\label{nhs}
\mathbb{J}_{12}\Pi^{\pm}|\lambda\rangle=(\lambda\pm 1)\Pi^{\pm}|
\lambda\rangle ~ 
\Rightarrow ~ \Pi^{\pm}|\lambda\rangle=\mu |\lambda\pm 1\rangle
\eea
where the normalization of state $\Pi^{\pm}|\lambda\rangle$ is fixed by 
the orthonormality of the helicity eigenstates and their Casimir 
eigenvalue. Hence by a repeated action of $\Pi^{\pm}$ we can construct an 
infinite set of linearly independent states with integer separated 
helicity values
\bea{l}\n
(\Pi^{\pm})^n|\lambda\rangle=\mu^n |\lambda\pm n\rangle~~,~~ 
n\in\mathbb{N}
\eea
All of these states belong in the same representation because they have 
the same $\mathcal{C}_2$ eigenvalue $\mu$
\bea{l}\n
\mathcal{C}_2~(\Pi^{\pm})^n|\lambda\rangle=\mu^2 ~ 
(\Pi^{\pm})^n|\lambda\rangle~~,~~n=0,1,2,...
\eea
Furthermore, by doing a full rotation of the state $|\lambda\rangle$ we 
get $e^{2i\pi\lambda}|\lambda\rangle$, thus
$\lambda$ is either $\pm N$ where $N$ is a non-negative integer (singled-
valued representation) or $\pm N/2$ where $N$ is a positive half odd 
integer (doubled-valued representation). The conclusion is
that the irreducible representation of E2 algebra (\ref{Pi}) are 
classified by a dimensionfull, continuous, parameter $\mu$ and for
$\mu\neq0$ they are infinite dimensional. The single-valued representation 
includes all integer helicity states and the doubled-valued
representation includes all the half odd integer helicity states. For the 
special case of $\mu=0$, the action of generators $\Pi^{\pm}$
becomes trivial and does not lead to new states. Therefore, the infinite 
size representation collapses to a dimension two representation with the 
states $|\lambda\rangle~,~|-\lambda\rangle$. This special case corresponds 
to the description of massless particles with a fixed helicity.

On the other hand, the definition of the angle states is:
\bea{l}\n\label{angle}
\Pi^{\pm}|\theta\rangle=\mu\hspace{0.4ex} e^{\pm 
i\theta}\hspace{0.2ex}|\theta\rangle~~,~~
\mathcal{C}_2|\theta\rangle=\mu^2|\theta\rangle
\eea
Such a state can be expanded in the complete basis of helicity states, 
hence we can write the ansatz
\bea{l}\n
|\theta\rangle=\sum_{\lambda}~f_{\lambda}(\theta)~|\lambda\rangle
\eea
where $f_{\lambda}(\theta)$ are the expansion coefficients, which due to 
(\ref{angle}) must satisfy
\bea{l}\n
f_{\lambda}(\theta)= e^{\pm i\theta}\hspace{0.2ex}
f_{\lambda\mp 1}(\theta)
\eea
This condition fixes uniquely the expansion coefficients, up to an overall 
normalization constant
\bea{l}\n
f_{\lambda}(\theta)\sim e^{- i\lambda\theta}
\eea
hence the angle states are the Fourier dual states to the helicity states
\bea{l}\n
|\theta\rangle\sim\sum_{\lambda}~e^{- i\lambda\theta}~|\lambda\rangle~.
\eea
It is straightforward to see that under the rotation 
$e^{-i\alpha\mathbb{J}_{12}}$ the angle state $|\theta\rangle$
will result to the state $|\theta+\alpha\rangle$, thus giving to 
$\mathbb{J}_{12}$ the interpretation of translations in the
$\theta$ sector.
%
% This is suggestive that after the introduction of a new auxiliary
% $\theta$ 
coordinate, the $\mathbb{J}_{12}$
%can be realized as $i\pa_{\theta}$. 
%This is the starting point of Wigner's argument in order to derive the 
covariant constraints
%on fields tha describe the continuous spin representation.
%
\subsection[4D N=1 supersymmetric extension of Poincare algebra]
{$4D,~\mathcal{N}=1$ super-Poincar\'{e} algebra}
\label{sCSR}
For the supersymmetric extension of the Poincar\'{e} algebra with only one 
supersymmetry
\footnote{We follow the discussions in \cite{GGRS,BK}
}
 , we add to
the list of Poincar\'{e} generators the four fermionic generators of 
supersymmetry $\mathbb{Q}_{\a}$ and $\bar{\mathbb{Q}}_{\ad}$.
Therefore, in addition to (\ref{Poincare}) we must consider the following:
\bea{l}\n
[\mathbb{J}_{mn},\mathbb{Q}_{\a}]=i(\s_{mn})_{\a}{}^{\b}\mathbb{Q}_{\b}
~~,~~
[\mathbb{J}_{mn},\bar{\mathbb{Q}}^{\ad}]=i(\bar{\s}_{mn})^{\ad}{}_{\bd}
\bar{\mathbb{Q}}^{\bd}~,\\
{}[\mathbb{P}_{m},\mathbb{Q}_{\a}]=0~~,~~
[\mathbb{P}_{m},\bar{\mathbb{Q}}_{\ad}]=0~,\\
\lbrace\mathbb{Q}_{\a},\mathbb{Q}_{\b}\rbrace=0~~,~~
\lbrace\bar{\mathbb{Q}}_{\ad},\bar{\mathbb{Q}}_{\bd}\rbrace=0~~,~~
\lbrace\mathbb{Q}_{\a},\bar{\mathbb{Q}}_{\ad}\rbrace=
-(\sigma^{m})_{\a\ad}\mathbb{P}_{m}~.
\eea
In this case the stabilizer subgroup must preserve both $\mathbb{P}_{m}$ 
and  $\mathbb{Q}_{\a}$ and is straightforward to show that it is generated 
by $\mathbb{P}_{m}$ and $\mathbb{Z}_{m}$, where
\bea{l}\n
\mathbb{Z}_{m}=\tfrac{1}{2}\varepsilon^{mnrs}\mathbb{J}_{nr}\mathbb{P}_{s}
+c(\sigma_{m})^{\ad\a}[\mathbb{Q}_{\a},\bar{\mathbb{Q}}_{\ad}]~.
\eea
This is the supersymmetric version of the Pauli-Lubanski vector. It is 
important to realize that supersymmetry not only appears in the second 
term but also in the first term, through the generator
of rotations. This can be seen by considering the superfield realization 
of the generator of rotations which takes the form 
$\mathcal{J}_{mn}=-ix_{[m}\pa_{n]}
+i\thd^{\bd}(\bar{\s}_{mn})^{\ad}{}_{\bd}\bar{\pa}_{\ad}
-i\th^{\b}(\s_{mn})_{\b}{}^{\a}\pa_{\a}-i\mathcal{M}_{mn}$. Notice that 
this is not the same as the non-supersymmetric case, because the
generator of rotations can also act on the fermionic directions of 
superspace. The parameter $c$ is a numerical coefficient and it's value 
differs between the massive and massless case. This is because in the 
massless case, half the supersymmetry generators become trivial 
(vanish) and that qualitatively changes the structure of the algebra. To 
see this once again we identify $\mathbb{P}_{m}$ with $p_{m}=(-E,0,0,E)$. 
For this case, the supersymmetry algebra takes the form
\bea{l}\n
\lbrace\mathbb{Q}_{\a},\bar{\mathbb{Q}}_{\ad}\rbrace=-
\begin{pmatrix}
2 & 0\\
0 & 0
\end{pmatrix}
E
\eea
and leads to
\bea{l}\n\label{killQ2}
\mathbb{Q}_{2}=0,~\bar{\mathbb{Q}}_{\dot{2}}=0~.
\eea 
With these constraints taken into account, the value of $c$ is $c=-1/8$ 
and the algebra of the massless, supersymmetric Pauli-Lubanski vector 
$\mathbb{Z}_m$ is:
\bea{l}\n
[\mathbb{Z}_m,\mathbb{P}_n]=0~~,~~[\mathbb{Z}_m,\mathbb{Q}_{\a}]=0~,\\
{}[\mathbb{Z}^m,\mathbb{Z}^n]=i\varepsilon^{mnrs}\mathbb{Z}_r\mathbb{P}_s
~~,~~\mathbb{Z}^m\mathbb{P}_m=0~,\\
{}[\mathbb{J}_{mn},\mathbb{Z}_{r}]=i\eta_{mr}\mathbb{Z}_{n}
-i\eta_{nr}\mathbb{Z}_{m}~.
\eea
An example of a qualitative difference between the above algebra and the 
corresponding algebra for massive particles is the commutativity of $
\mathbb{Z}_m$ with $\mathbb{Q}_{\a}$. For massive representations
this is no longer true and there is a non-trivial right hand side in 
$[\mathbb{Z}_m,\mathbb{Q}_{\a}]$.

The orthogonality of $\mathbb{Z}_{m}$ with $\mathbb{P}_{m}$ fixes its 
structure to be the same as in the non-supersymmetric case:
\bea{l}\n\label{Z}
\mathbb{Z}_{m}=p_{m} Z +T_{m}
\eea
where 
$Z=\mathbb{J}_{12}-\tfrac{1}{8E}(\bar{\sigma}_{3})^{\ad\a}[\mathbb{Q}_{\a}
,\bar{\mathbb{Q}}_{\ad}]$
%=\mathbb{J}_{12}-\tfrac{1}{4}-\tfrac{1}{4E}\mathbb{Q}_{1}\bar{\mathbb{Q}}
%_{\dot{1}}$ 
and $T_{m}$ is the transverse to $p_m$ vector as $\Pi_{m}$ in (\ref{W}),
$T_{m}=(0,T_1,T_2,0)$ with
$T_1=E(\mathbb{J}_{23}+\mathbb{J}_{20})
-\tfrac{1}{8E}(\bar{\sigma}_{1})^{\ad\a}
[\mathbb{Q}_{\a},\bar{\mathbb{Q}}_{\ad}],~
T_2=-E(\mathbb{J}_{13}+\mathbb{J}_{10})
-\tfrac{1}{8E}(\bar{\sigma}_{2})^{\ad\a}
[\mathbb{Q}_{\a},\bar{\mathbb{Q}}_{\ad}]$.
Using (\ref{killQ2}) the 
above expressions can be simplified
\bea{l}\n
Z=\mathbb{J}_{12}-\tfrac{1}{8E}[\mathbb{Q}_{1},\bar{\mathbb{Q}}_{\dot{1}}]
~,\\
T_1=E(\mathbb{J}_{23}+\mathbb{J}_{20})~~,~~T_2=-E(\mathbb{J}_{13}+
\mathbb{J}_{10})~.
\eea
and their algebra is
\bea{l}\n\label{T}
[T_1,T_2]=0~~,~~i[T_1,Z]=T_2~~,~~i[T_2,Z]=-T_1~.
\eea
%In the usual discussion of supersymmetric, massless representations (see
%\cite{BK}) the case of continuous spin representations is not considered
%and both $T_{1}$ and $T_{2}$ are set to zero by hand. However because
%the above algebra is identical to the non-supersymmetric (\ref{Pi}) one,
%the entire discussion for continuous spin representations
%of the Poincar\'{e} algebra can be applied as is to the super-Poincar\'{e}
%algebra by replacing $W$ and $\Pi_{i}$ with $Z$ and $T_i$ respectively 
%plus the additional conditions (\ref{killQ2}).
 %
Usually when
describing the supersymmetric, massless representations (see e.g.
\cite {GGRS}, \cite{BK}) the case of continuous spin representations
is considered as non-interesting or unworthy since it had no
relation to supersymmetric generalization of conventional field
theories. This is a main reason why such a case has not been
discussed in details. Therefore both $T_{1}$ and $T_{2}$ are set to
zero by hand like in non-supersymmetric case. However because the
above algebra is identical to the non-supersymmetric (\ref{Pi}) one,
the entire discussion for continuous spin representations of the
Poincar\'{e} algebra can be applied as is to the super-Poincar\'{e}
algebra by replacing $W$ and $\Pi_{i}$ with $Z$ and $T_i$
respectively plus the additional conditions (\ref{killQ2}).
%The last two  can be expressed covariantly in the following way:
%\bea{l}\n\label{killQ2}
%(\bar{\sigma}^m)^{\ad\a}p_m\mathbb{Q}_{\a}=0~~,~~(\bar{\sigma}^m)^{\ad\a}
%p_m\bar{\mathbb{Q}}_{\ad}=0~.
%\eea
Therefore the definition of supersymmetric
continuous spin representations with label $\mu$ is:
\bea{l}\n\label{superangle}
\mathcal{C}_2|\mu,\varphi\rangle=\mu^2|\mu,\varphi\rangle~~,~~
T^{\pm}|\mu,\varphi\rangle=\mu\hspace{0.4ex} e^{\pm 
i\varphi}\hspace{0.2ex}|\mu,\varphi\rangle~~,~~
%(\bar{\sigma}^m)^{\ad\a}p_m\mathbb{Q}_{\a}|\varphi\rangle=0~~,~~
%(\bar{\sigma}^m)^{\ad\a}p_m\bar{\mathbb{Q}}_{\ad}|\varphi\rangle=0
\mathbb{Q}_{2}|\mu,\varphi\rangle=0~~,~~
\bar{\mathbb{Q}}_{\dot{2}}|\mu,\varphi\rangle=0
\eea
where $T^{\pm}=T_1\pm i T_2$ and $\mathcal{C}_2=T^{+}~T^{-}$.
%
%%%%%%%%%%%%%%%%%%%%%%%%%%%%%%%%%%%%%%%%%%%%%%
\section{Superspace realization of supersymmetric continuous spin 
representations}
\label{sec3}
%%%%%%%%%%%%%%%%%%%%%%%%%%%%%%%%%%%%%%%%%%%%%%
The objective of this paper is to find a %superfield decription
$4D,~\mathcal{N}=1$ Minkowski, superspace realization of sCSR in order to 
make the connection under supersymmetry between the singled-valued and the 
double-valued CSR, manifest. Therefore, the use of superfields is a natural 
choice and the question is to find the appropriate superfield and the
necessary set of differential constraints required for the description of 
sCSR. These constraints have to be covariant under supersymmetry, so their 
nature does not change under a supersymmetry transformations. Hence, the 
constraints must be formulated in terms of the supersymmetry covariant 
derivatives $\D_{\a}$ and $\Dd_{\ad}$. Therefore, we must express the 
various objects that participate in the  discussion of section 2.2
in terms of the spinorial covariant derivatives and then impose the 
various diagonalization conditions.
\subsection{From Hilbert space to superspace}
As mentioned previously, once we consider the (super)field description of 
the various representations the abstract generators will be replaced by 
the familiar differential operators that describe the group action
in the space of (super)fields:
\bea{l}\n
\mathbb{J}_{mn}~\to~\mathcal{J}_{mn}=-ix_{[m}\pa_{n]}
+i\thd^{\bd}(\bar{\s}_{mn})^{\ad}{}_{\bd}\bar{\pa}_{\ad}-i\th^{\b}
(\s_{mn})_{\b}{}^{\a}\pa_{\a}-i\mathcal{M}_{mn}~,\\
\mathbb{Q}_{\a}~\to~\mathit{Q}_{\a}=i\pa_{\a}+\tfrac{1}{2}\thd^{\ad}
(\s^m)_{\a\ad}\pa_{m}~,\\
\bar{\mathbb{Q}}_{\ad}~\to~{\mathit{\bar{Q}}}_{\ad}=i\bar{\pa}_{\ad}
+\tfrac{1}{2}\th^{\a}(\s^m)_{\a\ad}\pa_{m}~\\
\mathbb{P}_{m}~\to~\mathcal{P}_{m}=-i\pa_{m}=p_m
\eea
The set of spinorial covariant derivatives with respect to supersymmetry 
are
\bea{l}\n
\D_{\a}=\pa_{\a}+\tfrac{i}{2}\thd^{\ad}(\s^m)_{\a\ad}\pa_{m}~~,~~
\Dd_{\ad}=\bar{\pa}_{\ad}+\tfrac{i}{2}\th^{\a}(\s^m)_{\a\ad}\pa_{m}~.
\eea
From the above, trivially the relation between $\mathit{Q}_{\a}$ and 
$\D_{a}$ can be written as:
\bea{l}\n\label{Q2D}
\mathit{Q}_{\a}+i\D_{\a}=2i\pa_{\a}~~,~~{\mathit{\bar{Q}}}_{\ad}
+i\Dd_{\ad}=2i\bar{\pa}_{\ad}
\eea
which can be used to convert between $\mathit{Q}$s and $\D$s. The 
constraints (\ref{killQ2})
can be written covariantly in the form 
$(\bar{\sigma}^m)^{\ad\a}p_m\mathbb{Q}_{\a}=0,~
(\bar{\sigma}^m)^{\ad\a}p_m\bar{\mathbb{Q}}_{\ad}=0$ and in superspace
they are translated to:
\be\label{con1}
(\bar{\sigma}^m)^{\ad\a}p_m\mathit{Q}_{\a}=0~\Rightarrow~
\begin{cases}
(\bar{\sigma}^m)^{\ad\a}p_m\pa_{\a}=0~,\\
(\bar{\sigma}^m)^{\ad\a}p_m\D_{\a}=0 ~\Rightarrow~[\D^2,\Dd_{\ad}]=0
\end{cases}
\ee
\be\label{con2}
(\bar{\sigma}^m)^{\ad\a}p_m\mathit{\bar{Q}}_{\ad}=0~\Rightarrow~
\begin{cases}
(\bar{\sigma}^m)^{\ad\a}p_m\bar{\pa}_{\ad}=0~,\\
(\bar{\sigma}^m)^{\ad\a}p_m\Dd_{\ad}=0 ~\Rightarrow~[\Dd^2,\D_{\a}]=0
\end{cases}
\ee
The expression for the supersymmetric Pauli-Lubanski vector 
$\mathbb{Z}_{m}$ is:
\bea{l}\n
\mathbb{Z}^{m}~\to 
~\mathit{Z}^{m}=-\tfrac{i}{2}\varepsilon^{mnrs}\mathcal{M}_{nr}p_s
+\tfrac{1}{8}(\bar{\s}^m)^{\ad\a}[\D_{\a},\Dd_{\ad}]~.
\eea
In the last one, it is interesting to observe how the $\theta$-derivatives 
($\pa_{\a}$ and $\bar{\pa}_{\ad}$) originating from (\ref{Q2D})
combined with their constraints (\ref{con1},\ref{con2}) cancel the theta 
dependent part of $\mathcal{J}_{mn}$ leaving only the usual
(internal) Poincar\'{e} part.
Therefore, according to the decomposition (\ref{Z}) we find the following 
for $Z$ and $T_{i}$
\bea{l}\n
Z=-i\mathcal{M}_{12}+\tfrac{1}{8E}(\bar{\s}_3)^{\ad\a}[\D_{\a},\Dd_{\ad}]
~,\\
T_1=-iE(\mathcal{M}_{23}+\mathcal{M}_{20})
+\tfrac{1}{8E}(\bar{\s}_1)^{\ad\a}[\D_{\a},\Dd_{\ad}]~,~\\
T_2=iE(\mathcal{M}_{13}+\mathcal{M}_{10})
+\tfrac{1}{8E}(\bar{\s}_2)^{\ad\a}[\D_{\a},\Dd_{\ad}]~.
\eea
Once again, we can use the constraints $(\bar{\sigma}^m)^{\ad\a}p_m\D_{\a}
=0=(\bar{\sigma}^m)^{\ad\a}p_m\Dd_{\ad}$ to simplify the above expressions 
(only the $\D_{1}$ and $\Dd_{\dot{1}}$ parts survive):
\bea{l}\n
Z=-i\mathcal{M}_{12}+\tfrac{1}{8E}[\D_{1},\Dd_{\dot{1}}]~~,~~
T_1=-iE(\mathcal{M}_{23}+\mathcal{M}_{20})~,~
T_2=iE(\mathcal{M}_{13}+\mathcal{M}_{10})~.
\eea
Notice that the $T_{i}$ found above do not seem to be aware of the 
presence of supersymmetry and match precisely the non-supersymmetry 
discussion. The only contribution of supersymmetry in the definition of 
sCSR seem to be the $\D$-constraints (\ref{con1},\ref{con2}).
\subsection{Superfield description of sCSR}
Looking back to the wavefunction description of CSR, there are two clues 
that provide some guidance. The first one is the infinite 
size of the representations with all integer separated helicities 
participating in the 
spectrum of the theory. That means that we can not describe CSR with 
a finite collection of tensor fields and one should consider the countable 
infinite set of increasing rank bosonic (or fermionic) tensor fields. The 
second clue is that the action of rotations on the angle states gives a 
shift in the angle parameter. This indicates that the intrinsic spin 
generator $\mathcal{M}_{mn}$ can be interpreted as the derivation with respect to an 
appropriate ``\emph{internal}'' coordinate not related to spacetime.
Both of these features suggest that one should introduce an auxiliary 
coordinate $\xi_{m}$ and consider the generating ``functions'' 
$\phi(\xi,x)$.
Morally, an expansion in terms of $\xi_m$ will generate an infinite list 
of spacetime fields with all possible ranks and the $\xi$-orbital angular 
momentum generator $\xi_{[m}\pi_{n]}$ \footnote{$\pi_{m}$ is the conjugate 
variable to $\xi_{m}$ such that $[\xi^{m},\pi_{n}]=i\delta^{m}_{n}$} will 
correspond 
to the intrinsic spin generator ($-i\mathcal{M}_{mn}$) and thus giving to 
all these fields the appropriate helicity value. The role of this 
auxiliary coordinate is to provide a mechanism in order to group the 
infinite set of components in to a multiplet with the correct book keeping 
for their helicities in order to match the spectrum of CSR. This approach 
turned out successful and provides the correct (Wigner's) covariant 
conditions for the field description of CSR.

All these features remain 
present in the case of sCSR, thus it is natural to follow a similar path.
For these reasons we consider an expand version of superspace by 
introducing the auxiliary coordinate $\xi_{m}$ such that the action
of internal spin generator $-i\mathcal{M}_{mn}$ on standard superspace 
superfield tensors is reproduced by the action of $-i\xi_{[m}
\tfrac{\pa}{\pa\xi^n]}$ on the extended superspace, rank zero (scalar)
superfield $\Phi(\xi,x,\th,\thd)$. Notice that the extension of superspace 
takes place only in the bosonic sector, in order to keep the same number 
of superchargers. Therefore $T_{i}$ can be written as:
\bea{l}\n
T_{1}=-i\xi_{2}~p^m\frac{\pa}{\pa\xi^m}+i\frac{\pa}{\pa\xi^2}~p^m\xi_m
~,~\\
T_{2}=i\xi_{1}~p^m\frac{\pa}{\pa\xi^m}-i\frac{\pa}{\pa\xi^1}~p^m\xi_m
\eea

% something about how this is identical to the nonsupersymmetric case and
% the choice between the dual constraints

The definition of sCSR is:
\bea{ll}\n\label{W1}
T_{i}\Phi(x,\xi,\th,\thd)\sim \mu\Phi(x,\xi,\th,\thd) ~&\Rightarrow ~
\begin{cases}
p^m\xi_{m}\Phi(x,\xi,\th,\thd)=0~,~\\
p^m\frac{\pa}{\pa\xi^m}\Phi(x,\xi,\th,\thd)=i\mu\Phi(x,\xi,\th,\thd)
\end{cases}\\
\mathcal{C}_{2}\Phi(x,\xi,\th,\thd)=\mu^2\Phi(x,\xi,\th,\thd)
~&\Rightarrow ~
\xi^m\xi_m\Phi(x,\xi,\th,\thd)=\Phi(x,\xi,\th,\thd)\n\label{W2}\\
\mathit{Q}_{2}\Phi(x,\xi,\th,\thd)=0~&\Rightarrow ~
[\D^2,\Dd_{\ad}]\Phi(x,\xi,\th,\thd)=0\n\label{susy1}\\
\mathit{\bar{Q}}_{\dot{2}}\Phi(x,\xi,\th,\thd)=0~&\Rightarrow ~
[\Dd^2,\D_{\a}]\Phi(x,\xi,\th,\thd)=0\n\label{susy2}
\eea
The first  three equations are Wigner's conditions for CSR. The last two
are the additional supersymmetric constraints in order to describe sCSR.
Equations (\ref{susy1},\ref{susy2}) are solved by either a chiral
superfield $\Phi$ ($\Dd_{\ad}\Phi=0$) with the equation of motion
$\D^2\Phi=0$ or by a complex linear superfield $\Sigma$ ($\Dd^2\Sigma=0$)
with the equation of motion $\D_{\a}\Sigma=0$.
%
%%%%%%%%%%%%%%%%%%%%%%%%%%%%%%%%%%%%%%%%%%%%%%
\section{Components discussion and the recovery of the
single and double valued CSR}
\label{components}
%%%%%%%%%%%%%%%%%%%%%%%%%%%%%%%%%%%%%%%%%%%%%%
For the case of the chiral superfield description, eq
(\ref{W1},\ref{W2},\ref{susy1},\ref{susy2}) take the form
\footnote{Convert the vector index to spinorial indices
$\xi^{\a\ad}=\tfrac{1}{2}(\bar{\s}^m)^{\ad\a}\xi_{m}$
and $\frac{\pa}{\pa\xi^{\a\ad}}=(\s^m)_{\a\ad}\frac{\pa}{\pa\xi^m}$}
:
\bea{l}
\xi^{\a\ad}\Dd_{\ad}\D_{\a}\Phi=0~,\n\\
\Dd^{\ad}\D^{\a}\frac{\pa}{\pa\xi^{\a\ad}}\Phi=-i\mu\Phi~,\n\\
\xi^{\a\ad}\xi_{\a\ad}\Phi=\tfrac{1}{2}\Phi~,\n\label{dist1}\\
\Dd_{\ad}\Phi=0~,\n\label{ch1}\\
\D^2\Phi=0~.\n\label{ch2}
\eea 
By projecting these superspace equations into equations for the component 
fields of $\Phi$ we find that the lowest component 
$\phi=\Phi|_{\th=0=\thd}$ describes the singled valued CSR
\bea{l}\n\label{phi}
\xi^{\a\ad}\pa_{\a\ad}\phi=0~,~
\pa^{\a\ad}\frac{\pa}{\pa\xi^{\a\ad}}\phi=-\mu\phi~,~
\xi^{\a\ad}\xi_{\a\ad}\phi=\tfrac{1}{2}\phi~,~
\Box\phi=0
\eea
and the lowest fermionic component $\psi_{\a}=\D_{\a}\Phi|_{\th=0=\thd}$
describes the doubled-valued CSR:
\bea{l}\n\label{psi}
\xi^{\a\ad}\pa_{\a\ad}\psi_{\b}=0~,~
\pa^{\a\ad}\frac{\pa}{\pa\xi^{\a\ad}}\psi_{\b}=-\mu\psi_{\b}~,~
\xi^{\a\ad}\xi_{\a\ad}\psi_{\b}=\tfrac{1}{2}\psi_{\b}~,~
\pa^{\a}{}_{\ad}\psi_{\a}=0
\eea

Similarly, the complex linear superfield description of sCSR takes the
form:
\bea{l}
\xi^{\a\ad}\D_{\a}\Dd_{\ad}\S=0~,\n\\
\D^{\a}\Dd^{\ad}\frac{\pa}{\pa\xi^{\a\ad}}\S=-i\mu\S~,\n\\
\xi^{\a\ad}\xi_{\a\ad}\S=\tfrac{1}{2}\S~,\n\label{dist2}\\
\Dd^2\S=0~,\n\label{s1}\\
\D_{\a}\S=0~.\n\label{s2}
\eea 
and it is straightforward to show that the components 
$\varphi=\S|_{\th=0=\thd}$ and $\lambda=\D_{\a}\bar{\S}|_{\th=0=\thd}$ satisfy 
the same (\ref{phi},\ref{psi}) conditions and thus provide a description 
of integer and half-integer CSR. Of course this alternative description of
sCSR exist due to the well-known duality between chiral and complex linear 
superfields which flips eq. (\ref{ch1},\ref{ch2}) with (\ref{s1},\ref{s2}).

An interesting observation is that due to (\ref{dist1},\ref{dist2})
the solutions must be searched in the space of distributions. This is a
characteristic property of CSR and sCSR. In addition, the coordinate $\xi_{\a\ad}$ can not be written as the product of two twistors because that will make it a lightlike coordinate ($\xi_{\a\ad}\neq\omega_{\a}\bar{\omega}_{\ad}$). However one can introduce two sets of twistors $\omega^{I}_{\a}$ with $I=1,2$. Then one can decompose $\xi_{\a\ad}$ in the following manner:
\bea{l}\n
\xi_{\a\ad}=\omega^{I}_{\a}\bar{\omega}^{J}_{\ad}\varepsilon_{IJ}
\eea
This decomposition makes contact with the description in \cite{rcsr22,rcsr23} where
two twistors $\pi_{\a}$ and $\rho_{a}$ and their conjugates where used
for the description of CSR. The correspondence is $\omega^{I}_{\a}=\lbrace \pi_{\a},\rho_{\a}\rbrace$ and thus it will relate the component fields found here with the ones used in the BRST description done in \cite{rcsr22}.
 
\section{Summary and discussion}
\label{outro}
%%%%%%%%%%%%%%%%%%%%%%%%%%%%%%%%%%%%%%%%%%%%%%
In this work we define the supersymmetric continuous spin representation 
(sCSR) (eq. \ref{superangle}). To find a superfield description of it we 
extended standard $4D, ~\mathcal{N}=1$ superspace with the addition of an
auxiliary, commuting, coordinate $\xi_{\a\ad}$ in order to construct 
generating superfunction that 
group together the countable infinite number of supersymmetric multiplets 
of increasing superhelicity that appear in the spectrum of sCSR. These are 
the supersymmetric extension of Wigner's wavefunctions  used for the 
description of CSR. We find two descriptions. The first is based on a
a chiral superfield $\Phi(\xi,x,\t,\thd)$ ($\Dd_{\ad}\Phi$=0) and the proposed set of 
covariant equations of motion it must satisfy is:
\bea{l}
\xi^{\a\ad}\Dd_{\ad}\D_{\a}\Phi=0~,\\
\Dd^{\ad}\D^{\a}\frac{\pa}{\pa\xi^{\a\ad}}\Phi=-i\mu\Phi~,\\
\xi^{\a\ad}\xi_{\a\ad}\Phi=\tfrac{1}{2}\Phi~,\\
\D^2\Phi=0~.
\eea 
The second description is dual to the first one and 
uses a complex linear superfield $\S(\xi,x,\t,\thd)$ ($\Dd^2\S=0$)
and it must satisfy the following equations:
\bea{l}
\xi^{\a\ad}\D_{\a}\Dd_{\ad}\S=0~,\\
\D^{\a}\Dd^{\ad}\frac{\pa}{\pa\xi^{\a\ad}}\S=-i\mu\S~,\\
\xi^{\a\ad}\xi_{\a\ad}\S=\tfrac{1}{2}\S~,\\
\D_{\a}\S=0~.
\eea 
The projected components $\phi(\xi,x)=\Phi|_{\th=0=\thd}$ or 
$\varphi(\xi,x)=\S|_{\th=0=\thd}$ are the 
Wigner's wavefunctions that describe the
single valued continuous spin representation (spans integer helicities)
Similarly the components $\psi_{\a}(\xi,x)=\D_{\a}\Phi|_{\th=0=\thd}$
or $\lambda(\xi,x)=\D_{\a}\bar{\S}|_{\th=0=\thd}$ are Wigner's 
wavefunctions for the description of the doubled valued continuous spin 
representation (spans half odd integer helicities). Additionally, there 
are auxiliary fields $F(\xi,x)=\D^2\Phi|_{\th=0=\thd}$ or $\rho_{\a}
(\xi,x)=\D_{\a}\S|_{\th=0=\thd}$ which vanish automatically by the 
equations of motion. Nevertheless, these fields appear in the supersymmetry transformation of the components 
$\phi(\xi,x), \psi_{\a}(\xi,x)$ or $\varphi(\xi,x), \lambda_{\a}(\xi,x)$
as dictated by the chiral or complex linear superfields respectively.
These supersymmetry transformations are the off-shell completion of the
supersymmetry transformations between the integer and half-integer CSR in \cite{rcsr23}.

%%%%%%%%%%%%%%%%%%%%%%%%%%%%%%%%%%%%%%%%%%%%%%%
%%%%%%%%%% Acknowledgement %%%%%%%%%%%%%%%%%%%%
{\bf Acknowledgments}\\[.1in] \indent
The research of I.\ L.\ B.\ was supported in 
parts by Russian Ministry of Education and Science, project No. 3.1386.2017.
He is also grateful to RFBR grant, project No. 18-02-00153 for
partial support. The research of S.\ J.\ G.\ and K.\ K.\ is supported by the 
endowment of the Ford Foundation Professorship of Physics at 
Brown University. Also this work was partially supported by the U.S. National 
Science Foundation grant PHY-1315155. 
%%%%%%%%%%%%%%%%%%%%%%%%%%%%%%%%%%%%
%%%%%%%%% Bibliography %%%%%%%%%%%%%%%%%%%%

\end{document}